\documentclass[letterpaper, 10pt, conference]{IEEEtran}
\IEEEoverridecommandlockouts
\usepackage{cite}
\usepackage{amsmath,amssymb,amsfonts}
\usepackage{algorithmic}
\usepackage{graphicx}
\usepackage{textcomp}
\usepackage{xcolor}
\usepackage{xcolor,colortbl}
\usepackage{threeparttable}
\usepackage{stfloats}
\usepackage{booktabs}
\usepackage{makecell}
\usepackage{array}
\usepackage{multirow}
\usepackage{stfloats} % 프리앰블에 추가

\definecolor{Gray}{gray}{0.85}
\def\BibTeX{{\rm B\kern-.05em{\sc i\kern-.025em b}\kern-.08em
    T\kern-.1667em\lower.7ex\hbox{E}\kern-.125emX}}
\begin{document}

\title{Robust 3DMA GNSS Positioning in Urban Areas Using Conservative Candidate Set Determination Based on GNSS Shadow and Reflection\\
\thanks{This work was supported in part by the National Research Foundation of Korea (NRF), funded by the Ministry of Science and ICT (MSIT), Republic of Korea, under Grant RS-2024-00358298;
in part by the Unmanned Vehicles Core Technology Research and Development Program through the NRF and the Unmanned Vehicle Advanced Research Center (UVARC), funded by MSIT, Republic of Korea, under Grant RS-2020-NR046546;
in part by MSIT, Republic of Korea, under the Information Technology Research Center (ITRC) support program, supervised by the Institute of Information \& Communications Technology Planning \& Evaluation (IITP), under Grant IITP-2024-RS-2024-00437494;
in part by Grant RS-2024-00407003 from the ``Development of Advanced Technology for Terrestrial Radionavigation System'' project, funded by the Ministry of Oceans and Fisheries, Republic of Korea;
in part by the Korea Aerospace Administration (KASA), under Grant RS-2022-NR067078;
and in part by the Korea Institute for Advancement of Technology (KIAT), funded by the Ministry of Trade, Industry and Resources (MOTIR), under Grant RS-2026-25585869 (HRD Program for Industrial Innovation).}
}

\makeatletter
\newcommand{\linebreakand}{%
  \end{@IEEEauthorhalign}
  \hfill\mbox{}\par
  \mbox{}\hfill\begin{@IEEEauthorhalign}
}
\makeatother

\author{\IEEEauthorblockN{Sanghyun Kim}
\IEEEauthorblockA{
\textit{School of Integrated Technology} \\
\textit{Yonsei University}\\
Incheon, Korea \\
sanghyun.kim@yonsei.ac.kr}
\and
\IEEEauthorblockN{Jiwon Seo${}^{*}$}
\IEEEauthorblockA{
\textit{School of Integrated Technology} \\
\textit{Yonsei University}\\
Incheon, Korea \\
jiwon.seo@yonsei.ac.kr}
{\small${}^{*}$ Corresponding author}
}

\maketitle

\begin{abstract}
In urban environments, global navigation satellite system (GNSS) positioning is often degraded by signal blockage and reflection caused by surrounding buildings. 
To address this problem, various three-dimensional mapping-aided (3DMA) GNSS techniques have been proposed.
Among them, zonotope shadow and reflection matching (ZSRM) has demonstrated effective positioning performance by determining the candidate receiver position set using predicted GNSS shadow and reflection regions.
However, ZSRM remains vulnerable to satellite signal misclassification, which may lead to positioning failure.
To address this limitation, this paper proposes a conservative candidate set determination strategy for robust ZSRM positioning.
Instead of using only the most probable signal reception condition from an artificial intelligence (AI)-based classifier, the proposed method combines the candidate regions associated with the two most probable classes.
This allows the candidate receiver position set to be determined more conservatively, reducing the risk of positioning failure under signal misclassification.
The proposed method was evaluated using real-world urban GNSS data.
Experimental results show that the proposed method improved the positioning success rate to 100\% for RF, GBDT, and SVM classifiers, compared with 8.2\%--30.1\% for the existing ZSRM algorithm using individual classifier outputs and 96.6\% for ZSRM with unanimous voting.
Although the proposed method increased the position bounds by retaining additional candidate regions, it produced smaller position bounds than ZSRM with unanimous voting.
These results demonstrate that the proposed method substantially enhances the robustness of ZSRM under satellite signal misclassification in urban GNSS environments.
\end{abstract}


\section{Introduction}

The global navigation satellite system (GNSS) has become a fundamental technology for intelligent transportation systems, autonomous vehicles, and urban mobility services \cite{Lee22:Urban, Ma20:Articial, Lee22:Optimal, Park25:Toward, Park26:LNSS}.
However, its positioning performance is significantly degraded in urban areas, where satellite signals are frequently blocked, reflected, or diffracted by surrounding buildings \cite{Bresler16:GNSS, Park24:CSAC}.
These harsh signal reception conditions reduce satellite visibility and introduce large pseudorange errors, resulting in unreliable positioning solutions \cite{MacGougan02}.
Therefore, improving the accuracy and robustness of GNSS positioning in dense urban environments remains an important research challenge.

To address this problem, three-dimensional mapping-aided (3DMA) GNSS techniques have been widely investigated \cite{Zhong22:Multi, Lee23:Nonlinear, Miura15:GPS, Kim24:Performance}.
These techniques use 3D city models to predict satellite visibility or signal propagation paths in urban environments.
Among them, shadow matching estimates the receiver position by comparing the predicted satellite visibility with the satellite visibility inferred from GNSS measurements \cite{Groves11:Shadow, Wang13:GNSS, Adjrad18:Intelligent}.
Previous studies have demonstrated its ability to improve positioning accuracy in various urban environments.
Nevertheless, conventional shadow matching relies on a discrete grid-based representation of the receiver position, making the positioning accuracy dependent on the predefined grid spacing \cite{Groves15:GNSS}.
Although a finer grid can improve positioning accuracy, it leads to higher computational cost.
To overcome this limitation, set-based positioning approaches have been introduced, where the receiver position is represented as a geometric set rather than a collection of discrete grid points.

Zonotope shadow matching (ZSM) is a representative set-based shadow matching technique that determines the receiver candidate position set using GNSS shadow regions, where line-of-sight (LOS) signals from a satellite are blocked by buildings \cite{Kim25:Set, Bhamidipati22:Set}.
In ZSM, each satellite signal is first classified as either LOS or non-line-of-sight (NLOS). 
Based on the classification result, the candidate receiver position set is determined as the region outside the computed GNSS shadow for LOS signals or inside the computed GNSS shadow for NLOS signals.

Zonotope shadow and reflection matching (ZSRM) extends ZSM by incorporating GNSS reflections, where direct and reflected signals can be received simultaneously \cite{Kim26:Zonotope}.
Unlike ZSM, ZSRM classifies satellite signals into three reception conditions: LOS-only, LOS+NLOS, and NLOS-only.
These conditions correspond to receiving only the direct signal, both direct and reflected signals, and only reflected signals, respectively.
Based on these conditions, ZSRM determines the candidate set using both shadow and reflection regions, which enables more effective reduction of the candidate position set and improved urban positioning accuracy.

Despite its improved positioning accuracy, the existing ZSRM algorithm can be vulnerable to satellite signal misclassification \cite{Neamati23:Risk, Kim25:Enhancing}.
In ZSRM, each satellite signal is classified as LOS-only, LOS+NLOS, or NLOS-only using an artificial intelligence (AI)-based classifier.
Based on the classification result, the candidate receiver position set is refined using the corresponding shadow and reflection constraints. 
However, if a satellite signal is misclassified, an incorrect geometric constraint may be applied. 
As a result, the true receiver position can be excluded from the candidate set, leading to positioning failure. 

To improve the positioning robustness of ZSRM, this paper proposes a conservative strategy for determining the candidate receiver position set.
Rather than determining the candidate receiver position set using only the most likely signal reception condition, the proposed strategy considers the feasible regions corresponding to the two most likely classes.
By applying geometric constraints more conservatively, this approach reduces the risk of excluding the true receiver position when satellite signals are misclassified.
The proposed method is validated using real-world GNSS data collected in an urban environment, demonstrating improved robustness against satellite signal misclassification.

%The remainder of this paper is organized as follows.
%Section~II provides the background of the ZSRM algorithm and defines the problem addressed in this study.
%Section~III presents the proposed conservative candidate position determination strategy.
%Section~IV describes the experimental setup used for performance evaluation.
%The experimental results are discussed in Section~V, and Section~VI concludes the paper. 

\section{Background}

\subsection{ZSRM Algorithm}

ZSRM is a set-based 3DMA GNSS positioning algorithm that estimates the receiver position by iteratively updating a candidate receiver position set \cite{Kim26:Zonotope}.
In ZSRM, the area of interest (AOI) is first defined as an initial candidate region that contains the possible receiver position.
For each satellite, GNSS shadow and reflection regions are computed using a 3D city model.
The GNSS shadow represents the region where the direct signal from the satellite is blocked by buildings, whereas the GNSS reflection represents the region where both direct and reflected signals from the satellite can be received simultaneously.

\begin{figure}[t]
    \centering
    \includegraphics[width=\linewidth]{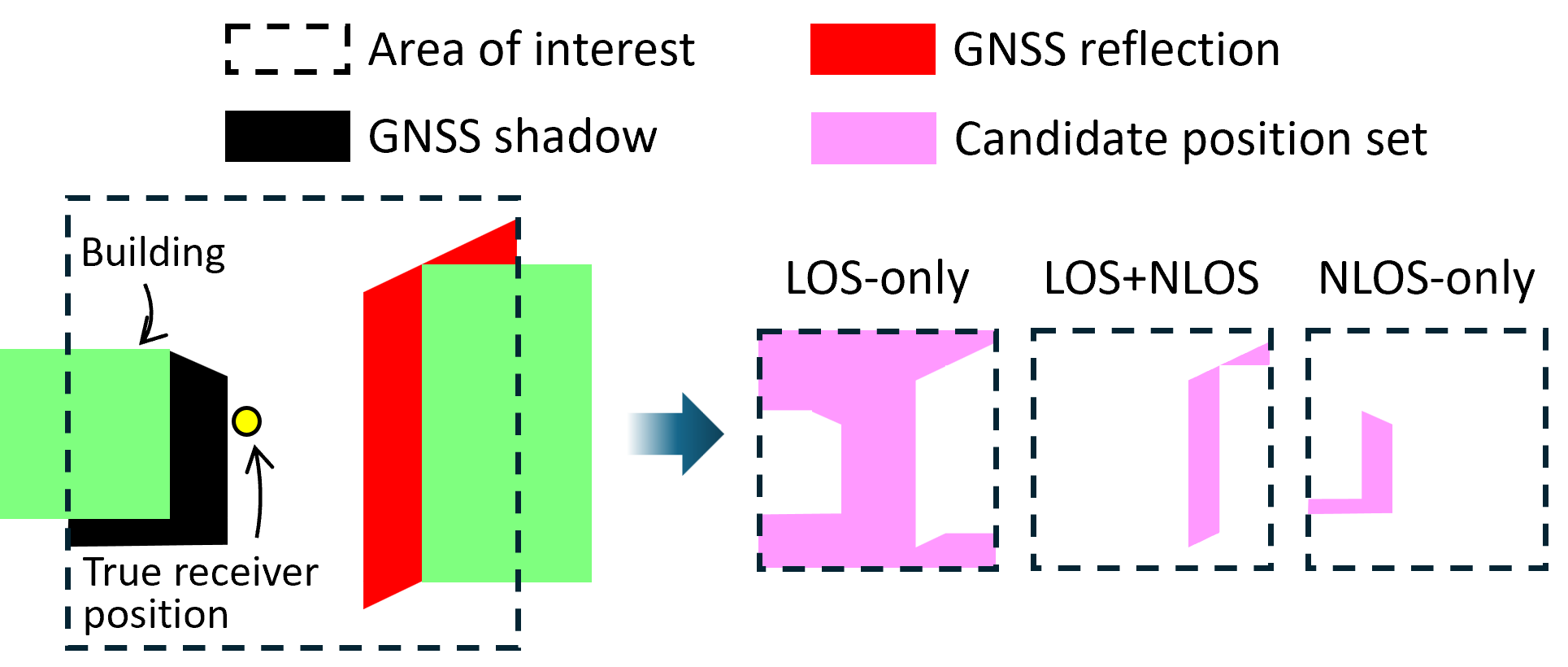}
    \caption{Candidate receiver position sets according to signal reception type in the ZSRM algorithm.}
    \label{fig:ZSRM_candidate_region}
\end{figure}

The computed GNSS shadow and reflection regions are then used to update the candidate receiver position set according to the signal reception type classified by an AI-based classifier using GNSS measurements \cite{Hsu17, Kim22:Machine, Kim23:Machine, Jeong24:Quantum}.
Fig.~\ref{fig:ZSRM_candidate_region} illustrates the candidate receiver position sets corresponding to the three signal reception types.
For LOS-only signals, the receiver is assumed to be located outside both the GNSS shadow and reflection regions.
For LOS+NLOS signals, the receiver is assumed to be located inside the GNSS reflection region.
For NLOS-only signals, the receiver is assumed to be located inside the GNSS shadow region.
Based on these class-dependent geometric conditions, ZSRM refines the candidate receiver position set by retaining only the region corresponding to the classified signal reception type.

This refinement process is sequentially applied to all available satellites.
For each satellite, the current candidate receiver position set is intersected with the region determined from the corresponding GNSS shadow or reflection constraint.
As more satellite-based constraints are applied, the candidate set is progressively reduced.
After all available satellites are processed, the final receiver position set is obtained, and the receiver position can be estimated from this final set, for example, by using its centroid.
Although this refinement rule enables ZSRM to constrain the receiver position geometrically, it also makes the resulting candidate set sensitive to errors in signal reception classification, as discussed in the following subsection.

\subsection{Problem Statement}

The existing ZSRM algorithm can effectively reduce the candidate receiver position set when the signal reception type of each satellite is correctly classified.
However, signal reception classification is not always perfect, and LOS-only, LOS+NLOS, and NLOS-only signals may be misclassified because they are not always clearly distinguishable from GNSS measurements \cite{Kim23:Single}. 
In this case, the candidate receiver position set may be refined using an incorrect GNSS shadow or reflection constraint.

\begin{figure}[t]
    \centering
    \includegraphics[width=\linewidth]{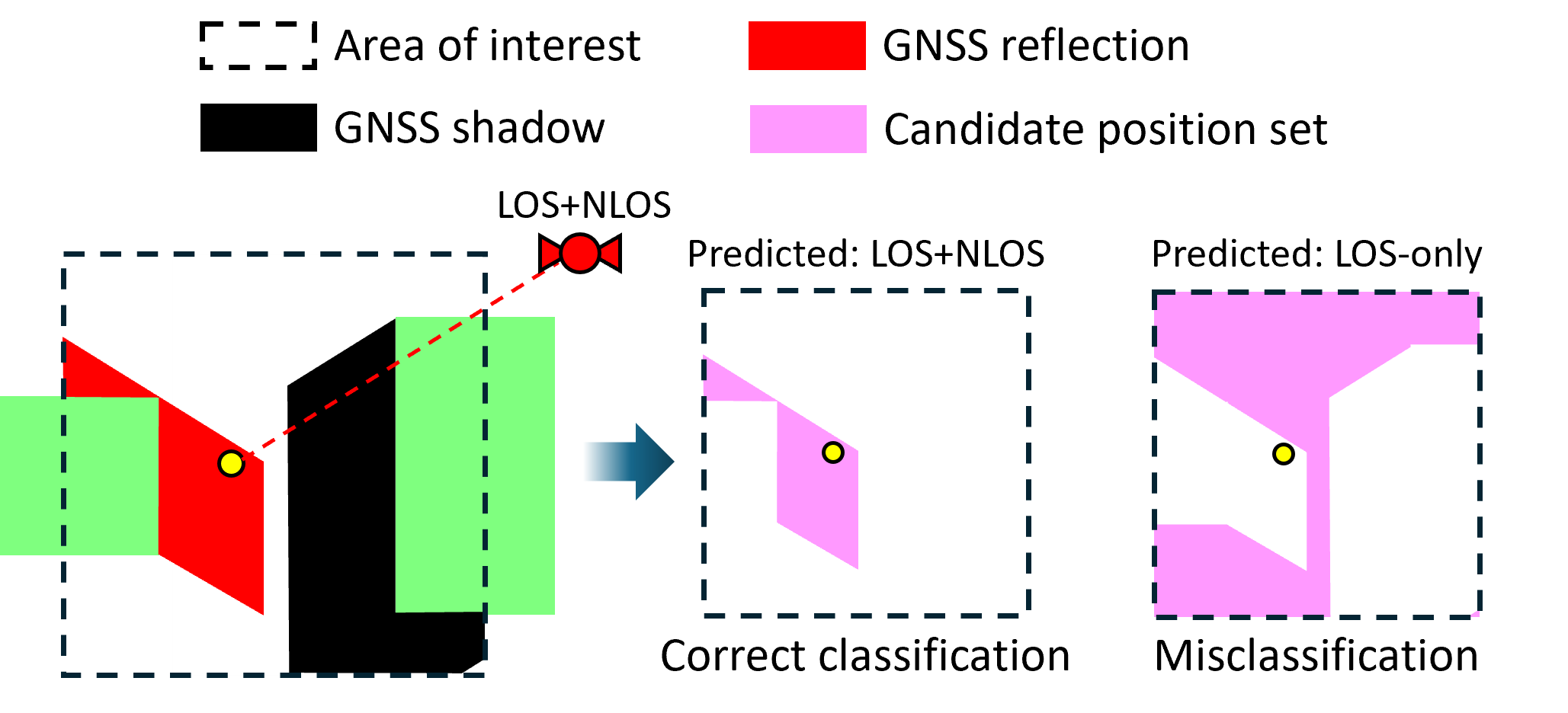}
    \caption{Effect of signal misclassification on the candidate position determination in ZSRM.}
    \label{fig:misclassification_problem}
\end{figure}

Fig.~\ref{fig:misclassification_problem} illustrates an example of the effect of signal misclassification on the candidate position determination in ZSRM.
In this example, the true receiver position is located inside the GNSS reflection region; therefore, the corresponding satellite signal should be classified as LOS+NLOS.
If the signal is correctly classified as LOS+NLOS, the candidate receiver position set is refined using the GNSS reflection region, and the true receiver position remains inside the refined candidate set.
However, if the same signal is misclassified as LOS-only, the candidate set is refined by excluding both the GNSS shadow and reflection regions.
As a result, the true receiver position can be removed from the candidate set.

This exclusion is particularly problematic because the ZSRM algorithm refines the candidate receiver position set sequentially over multiple satellites.
Once the true receiver position is excluded by an incorrect constraint, it cannot be recovered in the subsequent refinement process.
Therefore, even a small number of misclassified satellite signals can lead to positioning failure.

In the previous study \cite{Kim26:Zonotope}, a unanimous voting strategy was applied to mitigate the effect of signal misclassification by using only satellite signals unanimously classified by three machine learning classifiers in the ZSRM positioning process.
This approach improved the robustness of ZSRM by reducing the use of potentially misclassified satellite signals.
However, selecting only unanimously classified signals reduces the number of available satellites, which can limit the refinement of the candidate receiver position set \cite{Kim25:Enhancing}.
Therefore, this study aims to reduce the risk of excluding the true receiver position under satellite signal misclassification while maintaining the use of available satellite signals.
To achieve this objective, this paper introduces a conservative candidate set determination strategy for ZSRM.

\section{Proposed Method}
To improve the robustness of ZSRM under satellite signal misclassification, this paper proposes a conservative strategy for determining the candidate receiver position set.
Instead of relying only on the most probable class in the output of an AI-based classifier, the proposed method considers the candidate regions associated with the two most probable classes.
These regions are combined to form a conservative feasible region, which is then used to update the candidate receiver position set for each satellite.

Fig.~\ref{fig:proposed} illustrates an example of the proposed conservative candidate receiver position set determination.
In this example, the true signal reception type is LOS+NLOS, whereas the most probable class predicted by the AI-based classifier is LOS-only.
In the existing ZSRM algorithm, the candidate set is determined using only the LOS-only candidate region, causing the true receiver position to be excluded.
In contrast, the proposed method combines the candidate regions corresponding to LOS-only and LOS+NLOS.
As a result, the GNSS reflection region is retained as part of the conservative feasible region, and the true receiver position remains inside the resulting candidate set.

\begin{figure}[t]
    \centering
    \includegraphics[width=0.95\linewidth]{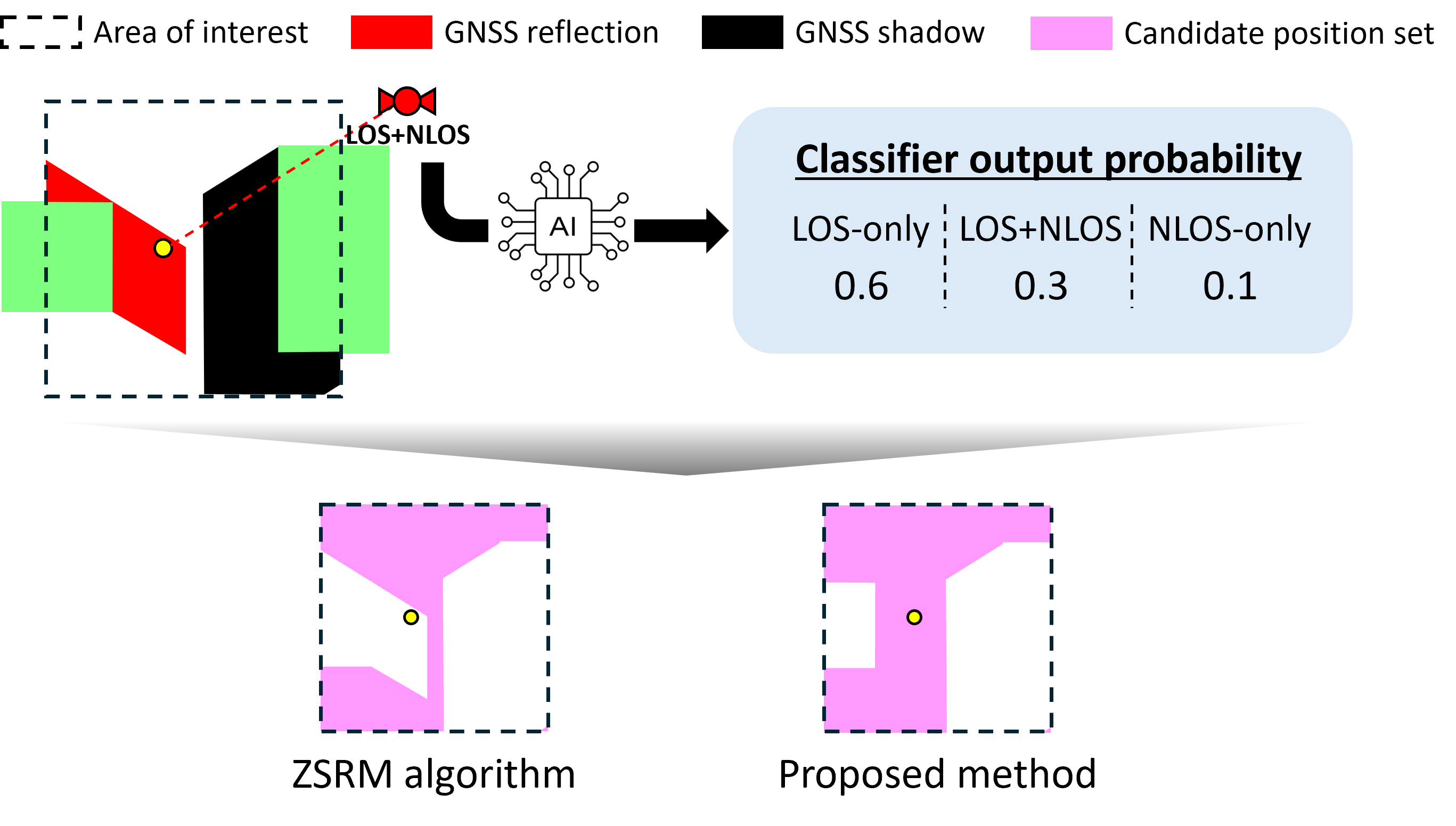}
    \caption{Illustration of candidate receiver position set determination in the proposed method.}
    \label{fig:proposed}
\end{figure}

The proposed candidate set determination process is applied sequentially to all available satellites in the same manner as in the existing ZSRM algorithm. 
Other procedures, including AOI definition and zonotope-based computation of GNSS shadow and reflection regions, remain unchanged from the existing ZSRM algorithm. 
The key difference is that the existing ZSRM uses a single candidate region for each satellite, whereas the proposed method constructs a conservative candidate region by combining the candidate regions associated with the two most probable classes. 
This reduces the risk of excluding the true receiver position under satellite signal misclassification.

\section{Experimental Setup}

This section describes the experimental setup used to evaluate the proposed conservative candidate set determination strategy.
The proposed method was evaluated using real-world GNSS data collected in an urban environment.
It was compared with the existing ZSRM algorithm using the same signal classification results to examine whether the proposed strategy reduces positioning failures caused by signal misclassification.

\subsection{GNSS Data Collection and Labeling}
The GNSS data used in this study were collected in Songdo, Incheon, South Korea.
Fig.~\ref{fig:dataset} shows the data collection routes and the corresponding dataset configuration used for training and testing.
The training dataset was constructed from GNSS data collected along multiple urban road segments, whereas the test dataset was collected along a separate road segment surrounded by high-rise buildings.
The target road segment was approximately 415 m long and 25 m wide.

A vehicle-based experiment was conducted to collect raw GNSS measurements along these routes.
The vehicle was equipped with an Antcom 3G1215RL-AA-XT-1 dual-polarized antenna and a NovAtel PwrPak7 GNSS receiver.
Raw GPS and Galileo measurements were recorded at 1 Hz during the experiment.

A reference trajectory was obtained using a NovAtel GPS-703-GGG antenna, a NovAtel SPAN-SE receiver, and a NovAtel UIMU-H58 inertial measurement unit.
The GNSS/INS measurements were post-processed using NovAtel Inertial Explorer in tightly coupled mode.
The resulting trajectory was used as the ground truth receiver position for signal labeling.

Each satellite signal was labeled as LOS-only, LOS+NLOS, or NLOS-only using the ground truth receiver position and a 3D city model of the test area provided by ONEGEO \cite{ONEGEO}.
For each satellite, GNSS shadow and reflection regions were computed based on the satellite geometry and surrounding buildings.
If the ground truth receiver position was located inside the GNSS shadow region, the signal was labeled as NLOS-only.
If the ground truth receiver position was located inside the GNSS reflection region, the signal was labeled as LOS+NLOS.
Otherwise, the signal was labeled as LOS-only.

Across the target road segment, the number of visible satellites per epoch ranged from 6 to 15.
On average, 6.4 LOS-only satellites, 2.9 LOS+NLOS satellites, and 1.8 NLOS-only satellites were observed per epoch.

\begin{figure}[t]
    \centering
    \includegraphics[width=0.75\linewidth]{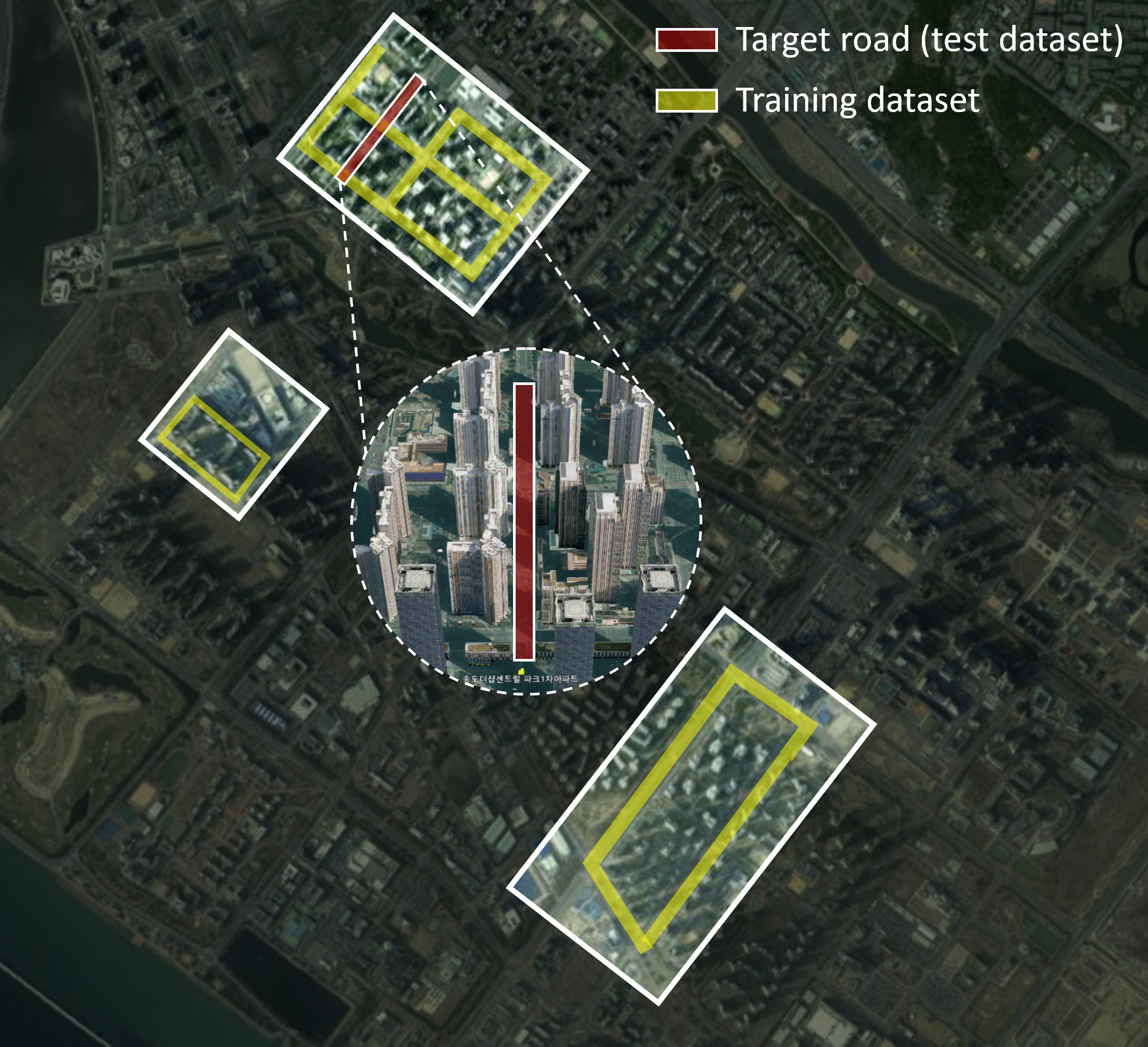}
    \caption{Overview of the data collection sites in Songdo, Incheon, South Korea.}
    \label{fig:dataset}
\end{figure}

\subsection{Signal Classification}
For both the existing ZSRM algorithm and the proposed method, satellite signals were classified using three machine learning classifiers: random forest (RF), gradient boosting decision tree (GBDT), and support vector machine (SVM).
The evaluation was conducted separately for RF, GBDT, and SVM to examine whether the proposed method provides consistent robustness improvements across different machine learning algorithms.
The input features consisted of the satellite elevation angle, carrier-to-noise density ratio ($C/N_0$), and pseudorange residual, which have been effectively used for GNSS signal classification in previous studies \cite{Xu24, Li23}.
For each classifier, the same classification output was used for both methods to ensure a fair comparison.

\subsection{Evaluation Metrics}
The proposed method was evaluated using four metrics: positioning success rate, receiver containment rate, cross-street position bound, and along-street position bound \cite{Kim25:Enhancing, Kim26:Zonotope}.
The positioning success rate was defined as the percentage of epochs in which a nonempty final candidate receiver position set was obtained after applying the constraints from all available satellites. 
The receiver containment rate was defined as the percentage of epochs in which the final candidate set contained the ground-truth receiver position.

The cross-street and along-street position bounds were used to evaluate the size of the final receiver position set.
These metrics were included because the proposed conservative candidate set determination strategy may preserve additional candidate regions, thereby reducing the degree of candidate set refinement.
Therefore, the evaluation focused on whether the proposed method can improve positioning robustness while maintaining reasonable position bounds.

\section{Results and Discussion}

\subsection{Signal Classification Results}
The signal classification performance of the three machine learning classifiers was first evaluated because the classifier output directly determines the candidate receiver position set in ZSRM. 
The top-1 classification accuracies were 63.8\% for RF, 66.8\% for GBDT, and 58.9\% for SVM. 
Among the three classifiers, GBDT achieved the highest top-1 accuracy, followed by RF and SVM. 
Given an average of 11.1 visible satellites per epoch, these accuracies indicate that approximately 3.69--4.56 satellite signals were misclassified per epoch on average.

In addition to the top-1 accuracy, the top-2 classification accuracy was evaluated because the proposed method uses the two most probable classes in the classifier output. The top-2 accuracies were 87.9\%, 89.5\%, and 84.2\% for RF, GBDT, and SVM, respectively. 
The higher top-2 accuracies indicate that even when the most probable class is incorrect, the correct signal reception type is often included as the second most probable class. 
This result supports the proposed conservative candidate set determination strategy, which preserves the candidate regions associated with the two most probable classes.

\subsection{Positioning Robustness}
The positioning robustness of the proposed method was evaluated by comparing its positioning success rate and receiver containment rate with those of the existing ZSRM algorithm.
Table~\ref{tab:success_rate} summarizes the results for the RF, GBDT, and SVM classifiers, as well as the unanimous voting strategy used in the previous study.
For the existing ZSRM algorithm, the positioning success rates were 25.3\%, 30.1\%, and 8.2\% for RF, GBDT, and SVM, respectively.
When unanimous voting was applied to the existing ZSRM algorithm, the positioning success rate increased to 96.6\%.
However, the proposed method further improved the positioning success rate across all classifiers, achieving 100\% for RF, GBDT, and SVM.
This indicates that the proposed method consistently maintains nonempty candidate sets without requiring the simultaneous use of multiple classifiers.

This improvement is consistent with the top-2 classification results discussed in the previous subsection.
The high top-2 classification accuracy indicates that the correct signal reception type was often included among the two most probable classes, even when the most probable class was incorrect.
In the existing ZSRM algorithm, an incorrect most probable class can impose an incorrect shadow or reflection constraint, which may remove the true receiver position from the candidate set.
In contrast, the proposed method determines the candidate receiver position set by combining the candidate regions corresponding to the two most probable classes.
Therefore, even when the most probable class is incorrect, the true receiver position can remain in the candidate set if the correct signal reception type is included as the second most probable class.
These results suggest that the proposed method reduces the risk of excluding the true receiver position under satellite signal misclassification.

In addition, the proposed method improved the receiver containment rates from 2.1\% to 15.1\% for RF, from 0.0\% to 45.2\% for GBDT, and from 0.0\% to 13.0\% for SVM.
However, these rates remained considerably lower than the positioning success rates because receiver containment requires the true receiver position to satisfy the geometric constraints of all processed satellites.
If even a single satellite constraint excludes the true receiver position, it cannot be recovered through subsequent intersections.
Among the three classifiers, GBDT achieved the highest receiver containment rate, which can be partly attributed to its higher top-2 classification accuracy.

\begin{table}
\centering
\caption{Comparison of positioning success and receiver containment rates}
\renewcommand{\arraystretch}{1.5}
\setlength{\tabcolsep}{3pt}
\begin{tabular}{
    >{\centering\arraybackslash}m{3.5cm}
    >{\centering\arraybackslash}m{2.3cm}
    >{\centering\arraybackslash}m{2.3cm}
}
\toprule
\textbf{Method} &
\makecell{\textbf{Positioning}\\\textbf{success rate}} &
\makecell{\textbf{Receiver}\\\textbf{containment rate}} \\
\midrule
ZSRM w/ RF & 25.3\% (37/146) & 2.1\% (3/146) \\
ZSRM w/ GBDT & 30.1\% (44/146) & 0.0\% (0/146) \\
ZSRM w/ SVM & 8.2\% (12/146) & 0.0\% (0/146) \\
ZSRM w/ unanimous voting & 96.6\% (141/146) & 19.9\% (29/146) \\
\rowcolor{gray!10}
Proposed method w/ RF & 100\% (146/146) & 15.1\% (22/146) \\
\rowcolor{gray!10}
Proposed method w/ GBDT & 100\% (146/146) & 45.2\% (66/146) \\
\rowcolor{gray!10}
Proposed method w/ SVM & 100\% (146/146) & 13.0\% (19/146) \\
\bottomrule
\end{tabular}
\label{tab:success_rate}
\end{table}

\subsection{Position Bound Analysis}

The position bounds were further evaluated to quantify the extent of the final receiver position sets obtained by each method.
In set-based positioning algorithms such as ZSRM, the final receiver position set represents the region in which the receiver may be located; therefore, its bound indicates how tightly the receiver position is constrained.
The position bounds were computed separately in the cross-street and along-street directions, as summarized in Tables~\ref{tab:position_bound} and~\ref{tab:position_bound2}.
To enable paired comparisons, the position bounds were evaluated over the epochs in which both compared methods successfully produced nonempty final receiver position sets.
Accordingly, Table~\ref{tab:position_bound} was evaluated using 37, 44, and 12 common successful epochs for RF, GBDT, and SVM, respectively, whereas Table~\ref{tab:position_bound2} was evaluated using 141 common successful epochs for the comparison with unanimous voting.

\begin{table}
\centering
\caption{Architecture-wise comparison of RMS cross-street and along-street position bounds over common successful epochs}
\renewcommand{\arraystretch}{1.8}
%\small
\begin{tabular}{
>{\centering\arraybackslash}m{3.5cm}
>{\centering\arraybackslash}m{2.0cm}
>{\centering\arraybackslash}m{2.0cm}
}
\toprule
\textbf{Method} & \makecell{\textbf{Cross-street}\\\textbf{position bound}} & \makecell{\textbf{Along-street}\\\textbf{position bound}} \\
\midrule
ZSRM w/ RF & 9.5 m & 10.5 m \\
\rowcolor{gray!10}
Proposed method w/ RF & 34.2 m & 43.8 m \\
\arrayrulecolor{gray!85}
\hline
\arrayrulecolor{black}
ZSRM w/ GBDT & 8.9 m & 7.8 m \\
\rowcolor{gray!10}
Proposed method w/ GBDT & 35.5 m & 45.3 m \\
\arrayrulecolor{gray!85}
\hline
\arrayrulecolor{black}ZSRM w/ SVM & 11.3 m & 8.1 m \\
\rowcolor{gray!10}
Proposed method w/ SVM & 33.5 m & 40.4 m \\
\bottomrule
\end{tabular}
\label{tab:position_bound}
\end{table}

\begin{table}
\centering
\caption{Comparison of RMS cross-street and along-street position bounds with unanimous voting over common successful epochs}
\renewcommand{\arraystretch}{1.8}
%\small
\begin{tabular}{>{\centering\arraybackslash}m{3.5cm} >{\centering\arraybackslash}m{2.0cm} >{\centering\arraybackslash}m{2.0cm}} 
\toprule
\textbf{Method} & \makecell{\textbf{Cross-street}\\\textbf{position bound}} & \makecell{\textbf{Along-street}\\\textbf{position bound}} \\
\midrule
ZSRM w/ unanimous voting & 45.8 m & 39.6 m \\
\rowcolor{gray!10}
\makecell{Proposed method w/ RF} & 30.7 m & 30.7 m \\
\rowcolor{gray!10}
\makecell{Proposed method w/ GBDT} & 29.9 m & 31.3 m \\
\rowcolor{gray!10}
\makecell{Proposed method w/ SVM} & 23.5 m & 25.1 m \\
\bottomrule
\end{tabular}
\label{tab:position_bound2}
\end{table}

As shown in Table~\ref{tab:position_bound}, the existing ZSRM algorithm using individual classifier outputs produced relatively small RMS cross-street and along-street position bounds. 
In contrast, the proposed method produced larger position bounds than the existing ZSRM algorithm for all three classifier architectures. 
This increase is expected because the proposed method preserves the candidate regions associated with the two most probable classes, rather than using only the most probable class. 
When considered together with the positioning success rates in Table~\ref{tab:success_rate}, these results indicate that the existing ZSRM algorithm tightly constrains the receiver position only for the epochs in which a valid final receiver position set is obtained, whereas the proposed method improves robustness to satellite signal misclassification at the cost of a larger final receiver position set.

A different trend is observed in Table~\ref{tab:position_bound2}, where the proposed method is compared with ZSRM using unanimous voting.
When applied separately to the RF, GBDT, and SVM outputs, the proposed method produced smaller RMS cross-street and along-street position bounds than ZSRM with unanimous voting.
In unanimous voting, satellite signals with non-unanimous classifier outputs are discarded, which can reduce the number of signals available for constraining the receiver position.
In contrast, the proposed method uses each classifier output individually and preserves the candidate regions associated with the two most probable classes.
Therefore, the proposed method can provide a more effective strategy for maintaining a compact final receiver position set while reducing the risk of excluding the true receiver position.

\section{Conclusion}
This paper proposed a conservative candidate set determination strategy to improve the positioning robustness of ZSRM under satellite signal misclassification.
Unlike the existing ZSRM algorithm, which determines the candidate receiver position set using only the most probable signal reception condition, the proposed method combines the candidate regions associated with the two most probable classes in the classifier output.
The results demonstrated that the proposed method achieved a 100\% positioning success rate for all three classifiers, compared with 8.2\%--30.1\% for the existing ZSRM algorithm using individual classifier outputs and 96.6\% for ZSRM with unanimous voting.
The proposed method also improved the receiver containment rate for all three classifiers, demonstrating its effectiveness in preserving the true receiver position under signal misclassification.
The position bounds were also analyzed to evaluate the size of the final receiver position set, showing that the proposed method produced larger cross-street and along-street bounds than the existing ZSRM algorithm because additional candidate regions were retained to reduce the risk of excluding the true receiver position.
Nevertheless, the proposed method maintained a 100\% positioning success rate while producing smaller position bounds than ZSRM with unanimous voting.
These results suggest that the proposed strategy improves ZSRM robustness under signal misclassification while maintaining more compact candidate sets than those obtained using unanimous voting.

\section*{Acknowledgment}

The authors acknowledge the use of a generative AI tool (ChatGPT, OpenAI) to assist with grammar and language editing during manuscript preparation.  
The AI tool was not used to generate technical content, ideas, data, or citations.  
All technical content, methodology, analysis, and conclusions are the sole work of the authors.

\bibliographystyle{IEEEtran}
\bibliography{mybibfile, IUS_publications}

\begin{thebibliography}{10}
\providecommand{\url}[1]{#1}
\csname url@samestyle\endcsname
\providecommand{\newblock}{\relax}
\providecommand{\bibinfo}[2]{#2}
\providecommand{\BIBentrySTDinterwordspacing}{\spaceskip=0pt\relax}
\providecommand{\BIBentryALTinterwordstretchfactor}{4}
\providecommand{\BIBentryALTinterwordspacing}{\spaceskip=\fontdimen2\font plus
\BIBentryALTinterwordstretchfactor\fontdimen3\font minus \fontdimen4\font\relax}
\providecommand{\BIBforeignlanguage}[2]{{%
\expandafter\ifx\csname l@#1\endcsname\relax
\typeout{** WARNING: IEEEtran.bst: No hyphenation pattern has been}%
\typeout{** loaded for the language `#1'. Using the pattern for}%
\typeout{** the default language instead.}%
\else
\language=\csname l@#1\endcsname
\fi
#2}}
\providecommand{\BIBdecl}{\relax}
\BIBdecl

\bibitem{Lee22:Urban}
H.~Lee, J.~Seo, and Z.~Kassas, ``Urban road safety prediction: A satellite navigation perspective,'' \emph{IEEE Intell. Transp. Syst. Mag.}, vol.~14, no.~6, pp. 94--106, Nov.-Dec. 2022.

\bibitem{Ma20:Articial}
Y.~Ma, Z.~Wang, H.~Yang, and L.~Yang, ``Artificial intelligence applications in the development of autonomous vehicles: a survey,'' \emph{IEEE/CAA J. Automatica Sinica}, vol.~7, no.~2, pp. 315--329, 2020.

\bibitem{Lee22:Optimal}
H.~Lee, S.~Pullen, J.~Lee, B.~Park, M.~Yoon, and J.~Seo, ``Optimal parameter inflation to enhance the availability of single-frequency {GBAS} for intelligent air transportation,'' \emph{IEEE Trans. Intell. Transp. Syst.}, vol.~23, no.~10, pp. 17\,801--17\,808, Oct. 2022.

\bibitem{Park25:Toward}
J.~Park, S.~Park, S.~Lee, J.~Seo, and E.~Kim, ``Toward high accuracy {DME} for alternative aircraft positioning: {SFOL} pulse transmission in high-power {DME},'' \emph{IEEE Trans. Aerosp. Electron. Syst.}, vol.~61, no.~5, pp. 15\,087--15\,097, 2025.

\bibitem{Park26:LNSS}
S.~Park and J.~Seo, ``{LNSS} onboard ephemeris-aided {Kalman} filter for clock synchronization in low visibility conditions,'' in \emph{Proc. ION PTTI}, Jan. 2026, pp. 132--141.

\bibitem{Bresler16:GNSS}
J.~Bresler, P.~Reisdorf, M.~Obst, and G.~Wanielik, ``{GNSS} positioning in non-line-of-sight context -- a survey,'' in \emph{Proc. IEEE ITSC}, 2016, pp. 1147--1154.

\bibitem{Park24:CSAC}
S.~Park and J.~H. Rhee, ``{CSAC} drift modeling considering {GPS} signal quality in the case of {GPS} signal unavailability,'' in \emph{Proc. ICCAS}, Oct.-Nov. 2024, pp. 187--192.

\bibitem{MacGougan02}
G.~MacGougan, G.~Lachapelle, R.~Klukas, K.~Siu, L.~Garin, J.~Shewfelt, and G.~Cox, ``Performance analysis of a stand-alone high-sensitivity receiver,'' \emph{GPS Solut.}, vol.~6, no.~3, pp. 179--195, 2002.

\bibitem{Zhong22:Multi}
Q.~Zhong and P.~D. Groves, ``Multi-epoch {3D}-mapping-aided positioning using {Bayesian} filtering techniques,'' \emph{Navig. J. Inst. Navig.}, vol.~69, no.~2, 2022.

\bibitem{Lee23:Nonlinear}
Y.~Lee, P.~Wang, and B.~Park, ``Nonlinear regression-based {GNSS} multipath dynamic map construction and its application in deep urban areas,'' \emph{IEEE Trans. Intell. Transp. Syst.}, vol.~24, no.~5, pp. 5082--5093, 2023.

\bibitem{Miura15:GPS}
S.~Miura, L.~T. Hsu, F.~Chen, and S.~Kamijo, ``{GPS} error correction with pseudorange evaluation using three-dimensional maps,'' \emph{IEEE Trans. Intell. Transp. Syst.}, vol.~16, no.~6, pp. 3104--3115, 2015.

\bibitem{Kim24:Performance}
S.~Kim and J.~Seo, ``Performance analysis of zonotope shadow matching algorithm according to various urban environments,'' \emph{J. Position. Navig. Timing}, vol.~13, no.~3, pp. 215--220, 2024.

\bibitem{Groves11:Shadow}
P.~D. Groves, ``Shadow matching: A new {GNSS} positioning technique for urban canyons,'' \emph{J. Navig.}, vol.~64, no.~3, pp. 417--430, 2011.

\bibitem{Wang13:GNSS}
L.~Wang, P.~D. Groves, and M.~K. Ziebart, ``{GNSS} shadow matching: Improving urban positioning accuracy using a {3D} city model with optimized visibility prediction scoring,'' \emph{Navig. J. Inst. Navig.}, vol.~60, no.~3, pp. 195--207, 2013.

\bibitem{Adjrad18:Intelligent}
M.~Adjrad and P.~D. Groves, ``Intelligent urban positioning: Integration of shadow matching with {3D}-mapping-aided {GNSS} ranging,'' \emph{J. Navig.}, vol.~71, no.~1, pp. 1--20, 2018.

\bibitem{Groves15:GNSS}
P.~D. Groves, L.~Wang, M.~Adjrad, and C.~Ellul, ``{GNSS} shadow matching: The challenges ahead,'' in \emph{Proc. ION GNSS+}, Sep. 2015, pp. 2421--2443.

\bibitem{Kim25:Set}
S.~Kim and J.~Seo, ``Set-based position ambiguity reduction method for zonotope shadow matching in urban areas using estimated multipath errors,'' in \emph{Proc. ION ITM}, Jan. 2025, pp. 1--10.

\bibitem{Bhamidipati22:Set}
S.~Bhamidipati, S.~Kousik, and G.~Gao, ``Set-valued shadow matching using zonotopes for {3D}-map-aided {GNSS} localization,'' \emph{Navig. J. Inst. Navig.}, vol.~69, no.~4, 2022.

\bibitem{Kim26:Zonotope}
S.~Kim and J.~Seo, ``Zonotope shadow and reflection matching: A novel {GNSS} reflection-based framework for enhanced positioning accuracy in urban areas,'' \emph{IEEE Trans. Intell. Transp. Syst.}, vol.~27, no.~5, pp. 5612--5628, May 2026.

\bibitem{Neamati23:Risk}
D.~Neamati, S.~Bhamidipati, and G.~Gao, ``Risk-aware autonomous localization in harsh urban environments with mosaic zonotope shadow matching,'' \emph{Artif. Intell.}, vol. 324, p. 104000, 2023.

\bibitem{Kim25:Enhancing}
S.~Kim and J.~Seo, ``Enhancing urban {GNSS} positioning reliability via conservative satellite selection using unanimous voting across multiple machine learning classifiers,'' in \emph{Proc. IEEE ITSC}, Nov. 2025, pp. 1504--1509.

\bibitem{Hsu17}
L.-T. Hsu, ``{GNSS} multipath detection using a machine learning approach,'' in \emph{Proc. IEEE ITSC}, Oct. 2017, pp. 1--6.

\bibitem{Kim22:Machine}
S.~Kim, J.~Byun, and K.~Park, ``Machine learning-based {GPS} multipath detection method using dual antennas,'' in \emph{Proc. ASCC}, May 2022, pp. 691--695.

\bibitem{Kim23:Machine}
S.~Kim and J.~Seo, ``Machine-learning-based classification of {GPS} signal reception conditions using a dual-polarized antenna in urban areas,'' in \emph{Proc. IEEE/ION PLANS}, Apr. 2023, pp. 113--118.

\bibitem{Jeong24:Quantum}
S.~Jeong, S.~Kim, and J.~Seo, ``Quantum support vector machine-based classification of {GPS} signal reception conditions,'' in \emph{Proc. QCE}, 2024, pp. 530--531.

\bibitem{Kim23:Single}
S.~Kim, S.~Park, and J.~Seo, ``Single antenna based {GPS} signal reception condition classification using machine learning approaches,'' \emph{J. Position. Navig. Timing}, vol.~12, no.~2, pp. 149--155, 2023.

\bibitem{ONEGEO}
\BIBentryALTinterwordspacing
{ONEGEO}. [Online]. Available: \url{https://onegeo.co/}
\BIBentrySTDinterwordspacing

\bibitem{Xu24}
P.~Xu, G.~Zhang, B.~Yang, and L.-T. Hsu, ``Machine learning in {GNSS} multipath/{NLOS} mitigation: Review and benchmark,'' \emph{IEEE Aerosp. Electron. Syst. Mag.}, vol.~39, no.~9, pp. 26--44, 2024.

\bibitem{Li23}
Z.~Li, K.~Zeng, L.~Wang, X.~Kan, R.~Yuan, and S.~Xie, ``{NLOS/LOS} classification by constructing indirect environment interaction from {GNSS} receiver measurements using a transformer-based deep learning model,'' in \emph{Proc. ION ITM}, 2023, pp. 859--870.

\end{thebibliography}

\end{document}